\pdfoutput=1
\newcommand*{\ATLASLATEXPATH}{latex/}
\documentclass[cernpreprint, texlive=2016, UKenglish]{\ATLASLATEXPATH atlasdoc}
\usepackage{\ATLASLATEXPATH atlaspackage}
\usepackage{\ATLASLATEXPATH atlasbiblatex}

\usepackage{\ATLASLATEXPATH atlasphysics}
\graphicspath{{logos/}{figures/}{fig/}}

\usepackage{EXOT-2017-01-PAPER-defs}
\usepackage{slashbox}

\AtlasTitle{Search for light resonances decaying to boosted quark pairs and produced in association with a photon or a jet in proton--proton collisions at $\sqrt{s}=13$~TeV with the ATLAS detector}

\newcommand{\Lumi}{36.1\xspace}

\AtlasAbstract{This Letter presents a search for new light resonances decaying to pairs of quarks and produced in association with a high-$p_{\textrm{T}}$ photon or jet.
The dataset consists  of proton--proton collisions with an integrated luminosity of \Lumi fb$^{-1}$ at a centre-of-mass energy of $\sqrt{s}=13$~TeV recorded by the ATLAS detector at the Large Hadron Collider. Resonance candidates are identified as massive large-radius jets with substructure consistent with a  
particle decaying into a quark pair. The  mass spectrum of the candidates is examined for local excesses above background. No evidence of a new resonance is observed in the data, which are used to exclude the production of a lepto-phobic axial-vector $Z^\prime$ boson.}

\author{The ATLAS Collaboration}

\AtlasRefCode{EXOT-2017-01}

\PreprintIdNumber{CERN-EP-2017-280}

\AtlasJournalRef{\PLB 788 (2019) 316}
\AtlasDOI{10.1016/j.physletb.2018.09.062}

\hypersetup{pdftitle={ATLAS document},pdfauthor={The ATLAS Collaboration}}

\begin{document}

\maketitle

%\tableofcontents

%-------------------------------------------------------------------------------
\section{Introduction}
\label{sec:intro}
%-------------------------------------------------------------------------------

Searches for resonance signals in the  invariant mass spectrum of hadrons are an essential part of the physics programme at the energy frontier.  Many theoretical models predict resonances~\cite{PhysRevD.34.1530,Langacker:2008yv,Salvioni:2009mt} with significant couplings to quarks and gluons, including resonances which also couple to dark-matter particles~\cite{An:2012ue,Rajaraman:2011wf,Goodman:2010ku,Goodman:2010yf}.  
At the Large Hadron Collider (LHC), the ability to discover or exclude such hadronic resonances has been extended into the \TeV\xspace range, although no evidence of statistically significant excesses has been seen~\cite{Aaboud:2017yvp,Sirunyan:2016iap}.

Sensitivity to light resonances is reduced by the immense background rates that would saturate the trigger and data acquisition systems.  The recording of collision data typically requires placing thresholds of several hundred GeV on the transverse momentum ($p_{\textrm{T}}^{\textrm{min}}$) of the jet used to trigger the event, which translates to approximate thresholds on mass of {\mbox{$m \approx 2p_{\textrm{T}}^{\textrm{min}}$}}. Consequently, recent searches for dijet resonances at the LHC have poor sensitivity for masses well below 1~\TeV. 
This limitation can be avoided by recording only a summary of the jet information needed for performing a resonance search in the dijet mass spectrum. This strategy is called ``data scouting'' in CMS~\cite{Khachatryan:2016ecr}, ``real-time analysis'' in LHCb~\cite{Aaij:2016rxn} and ``trigger-object-level analysis'' in ATLAS~\cite{Aaboud:2018fzt}, and has set limits for resonance masses in the range 500--800~\gev~\cite{Khachatryan:2016ecr}. 

\newcommand{\AtlasCoordFootnote}{%
ATLAS uses a right-handed coordinate system with its origin at the nominal interaction point (IP)
in the centre of the detector and the $z$-axis along the beam pipe.
The $x$-axis points from the IP to the centre of the LHC ring,
and the $y$-axis points upwards.
Cylindrical coordinates $(r,\phi)$ are used in the transverse plane, 
$\phi$ being the azimuthal angle around the $z$-axis.
The pseudorapidity is defined in terms of the polar angle $\theta$ as $\eta = -\ln \tan(\theta/2)$.
It is equivalent to the rapidity for massless particles. 
Angular distance is measured in units of $\Delta R \equiv \sqrt{(\Delta\eta)^{2} + (\Delta\phi)^{2}}$.}

In this Letter, a search using an alternative approach~\cite{An:2012ue,Shimmin:2016vlc} is performed, in order to cover even lower resonance masses. The trigger threshold limitations are reduced by examining data where the light resonance is boosted in the transverse direction\footnote{\AtlasCoordFootnote\xspace} via recoil from high transverse momentum ($p_{\textrm{T}}$) initial-state radiation (ISR) of a photon or jet.
Requiring a hard ISR object in the final state comes at the cost of reduced signal production rates, but allows highly efficient triggering at masses much lower than 
when triggering directly on the resonance decay products. 

The search is performed for 
resonance masses from 100~\GeV\xspace to 220~\GeV, a range in which the resonance is boosted and its decay products are collimated, such that the resonance mass can be calculated from the mass of a large-radius jet. 
The dominant background processes are multijet production in the jet channel and photons produced in association with jets in the photon channel, both characterised by non-resonant jets initiated predominantly by single gluons or light-flavour quarks. The $Z'$ signal models considered decay to quark-antiquark pairs. This difference in the dominant jet production mechanism between the signal and the leading backgrounds means that, in the boosted regime considered in this Letter, the use of jet substructure methods strongly suppresses the background, making it a crucial component for the search sensitivity.
In addition, current datasets are the largest collected, allowing the sensitivity to rare processes to be extended beyond that of earlier studies.

Recently, CMS reported results of applying a similar technique~\cite{Sirunyan:2017dnz,Sirunyan:2017nvi} to exclude a light $Z'$ boson with Standard Model (SM) coupling values ($g_{q}$) exceeding 0.1 to 0.25 in the mass range 50--300~\gev. With respect to those results, this Letter also exploits the channel with the ISR photon.

%-------------------------------------------------------------------------------
\section{ATLAS detector}
\label{sec:detector}
%-------------------------------------------------------------------------------

The ATLAS experiment~\cite{PERF-2007-01} at the LHC is a multi-purpose particle detector with a forward-backward symmetric cylindrical geometry with layers of tracking, calorimeter, and muon detectors over nearly the entire solid angle around the proton--proton ($pp$) collision point. 
The inner detector (ID) consists of a high-granularity silicon pixel detector,
including an insertable B-layer~\cite{ATLAS-TDR-19}, and a silicon microstrip tracker, together providing precision tracking in the pseudorapidity range $|\eta| < 2.5$. Complementary, a transition radiation tracker provides tracking and electron identification information for $|\eta| < 2.0$. The ID is surrounded by a 2 T superconducting solenoid.
Lead/liquid-argon (LAr) sampling calorimeters provide electromagnetic (EM) energy measurements with high granularity, covering the region $|\eta| < 3.2$. 
A hadron (steel/scintillator-tile) calorimeter covers the central pseudorapidity range ($|\eta| < 1.7$).
The end-cap and forward regions are instrumented with copper/LAr calorimeters ($1.7 < |\eta| < 3.2$) and LAr calorimeters with copper and tungsten absorbers, providing EM and hadronic energy measurements covering the region $|\eta| \le 4.9$. 
The muon spectrometer consists of precision tracking chambers covering the region $|\eta| \le 2.7$.
The first-level trigger is implemented in hardware and uses a subset of the detector information to reduce the accepted rate to 100 kHz. This hardware trigger~\cite{Aaboud:2016leb} is followed by a software-based trigger that reduces the rate of recorded events to 1 kHz.

%-------------------------------------------------------------------------------
\section{Data and simulation samples}
\label{sec:samples}
%-------------------------------------------------------------------------------

The data were collected in $pp$ collisions at $\sqrt{s} = 13$~\TeV\xspace during 2015 and 2016.  Collision events are recorded with two triggers. The first selects events with at least one photon candidate that has an online transverse energy $E_{\textrm{T}} > 140$~\GeV\xspace and passes the ``loose'' identification requirements based on the shower shapes in the EM and hadronic calorimeters~\cite{Aaboud:2016leb}.
The photon trigger reaches its maximum efficiency for $E_{\textrm{T}} > 155$~\GeV.
The second trigger selects events with at least one jet candidate with online $E_\mathrm{T}$ > 380 GeV formed from clusters of energy deposits in the calorimeters~\cite{Aad:2016upy} by the anti-$k_t$ algorithm~\cite{antikT,Cacciari:2006} with radius parameter $R= 0.4$, implemented in the {\textsc FastJet} package~\cite{Cacciari:2011ma}. The jet trigger reaches its maximum efficiency for $p_{\textrm{T}} > 420$~\GeV.
Only data satisfying beam, detector and data-quality criteria are considered~\cite{ATL-DAPR-PROC-2017-001}. The data used correspond to an integrated luminosity of \Lumi fb$^{-1}$. 

Samples of simulated events are used to characterise the hypothetical resonances as well as to study the kinematic distributions of background processes. These samples are not used to estimate the background contributions, except when validating the data-driven background estimate (described in Section~\ref{sec:background}).

Background samples were simulated using the \sherpa~2.1.1 event generator~\cite{Gleisberg:2008ta}.
Processes containing a photon with associated jets were  
generated in several bins of photon $p_{\textrm{T}}$. 
The matrix elements were calculated at leading order (LO) with up to three partons for photon $p_{\textrm{T}}<70$~\GeV\xspace or four partons for higher photon $p_{\textrm{T}}$. 
Multijet background samples were generated at LO in several bins of
leading-jet $p_{\textrm{T}}$. 
Samples of $W$+jets, $Z$+jets, $W$+$\gamma$ and $Z$+$\gamma$ events with hadronic decays of the vector-bosons were simulated in bins of $W/Z$-boson $p_{\textrm{T}}$. 
Matrix elements were calculated at LO with up to four partons for the $W/Z$+jets samples and up to three partons for $W/Z$+$\gamma$ samples. The cross sections 
were corrected at next-to-leading order (NLO) using $K$-factors derived from corresponding samples with leptonic vector-boson decays generated at NLO using \sherpa~2.1.1~\cite{Gleisberg:2008ta}, with matrix elements calculated for up to two partons at NLO and four partons at LO using Comix~\cite{Gleisberg:2008fv} and OpenLoops~\cite{Cascioli:2011va}. 
All the above LO background samples were merged with the \sherpa parton shower~\cite{Schumann:2007mg} 
using the {\scshape ME+PS@LO} prescription~\cite{Hoeche:2009rj}. 
The {\scshape CT10} set of parton distribution functions (PDFs)~\cite{Lai:2010vv} were used in conjunction with the dedicated parton shower tuning developed by the \sherpa authors. 
For the NLO leptonic vector-boson samples utilised to calculate $K$-factors, the {\scshape ME+PS@NLO} prescription~\cite{Hoeche:2009rj} and the {\scshape CT10nlo} PDF set are used.

As a benchmark signal, samples with a $Z^\prime$ resonance with only hadronic couplings were generated as in Refs.~\cite{Dobrescu:2013coa,LHCDMF:2015,Aad2016}. 
This $Z^\prime$ has axial-vector couplings to quarks. The coupling of the $Z^\prime$ 
to quarks, $g_{q}$, is set to be universal in quark flavour and equal to 0.5. 
The corresponding total width $\Gamma_{Z'}$ is negligible 
compared to the experimental resolution, which is about 10\% of the boson mass.
A set of samples was generated with $m_{Z^\prime}$ between 100 and 220~\GeV, in 30~\GeV~steps. 
A linear and parameterised 
interpolation was performed in 
10~\GeV~steps in between the generated mass points. 
The samples were produced with $g_{q}=0.5$, 
using the {\scshape MadGraph\_aMC@NLO} generator~\cite{Alwall:2014hca} with the NNPDF2.3 LO PDF~\cite{Ball:2012cx} 
and the A14 set of tuned parameters (tune)~\cite{A14tune}. Parton showers were produced in \pythia~8.186~\cite{pythia8}.
Interference of this benchmark model with the Standard Model $Z$ boson is assumed to be negligible.
For efficient population of the kinematic phase space, a photon (jet) with $p_\mathrm{T} \geq$ 100 GeV (350 GeV) was required in the generation phase.

The response of the detector to particles was modelled with a full ATLAS detector simulation~\cite{Aad:2010ah} based on $\textsc{Geant}4$~\cite{Agostinelli:2002hh}. 
All simulated events were overlaid with additional $pp$ interactions (pile-up) 
simulated with the soft strong-interaction processes of \pythia 8.186~\cite{pythia8} using the A2 tune~\cite{ATL-PHYS-PUB-2011-014} and the MSTW2008LO PDF set~\cite{Sherstnev:2007nd}. The simulated events were reconstructed in the same way as the data, and were reweighted such that the distribution of the expected number of $pp$ interactions per bunch crossing matches that seen in data.
 
%-------------------------------------------------------------------------------
\section{Event reconstruction and selection}
\label{sec:objects}
%-------------------------------------------------------------------------------

Events are required to have a reconstructed primary vertex, defined as a vertex with at least two reconstructed tracks
with $p_{\textrm{T}}>400$~MeV~each 
and with the largest sum of track $p_{\textrm{T}}^2$.

Photons are reconstructed from clusters of energy deposits in the electromagnetic calorimeter.  
The photon energy scale is corrected using
events with $Z \rightarrow e^+e^-$ decays in data~\cite{ATL-PHYS-PUB-2016-015}. 
Identification requirements are applied to reduce the contamination from $\pi^0$ or other neutral hadrons decaying into photons. The photon identification is based on the profile of the energy deposits in the first and second layers of the electromagnetic calorimeter.  Photons used in the event selection must satisfy the ``tight'' identification and isolation criteria defined in Ref.~\cite{ATL-PHYS-PUB-2016-014}, 
and must have
$|\eta| < 2.37$, excluding the EM calorimeter's barrel/end-cap transition region of $1.37 < |\eta| < 1.52$. 
The efficiency of the photon selection is roughly 95\% for photons with $E_{\textrm{T}}>150$~\GeV.

Two non-exclusive categories of jet candidates are built from clusters of energy deposits in the calorimeters~\cite{Aad:2016upy} and are distinguished by the radius parameter used in the anti-$k_t$ algorithm.  Jets with a radius parameter $R = 1.0$  are referred to as {\textit{large-R}} jets, denoted by $J$ and required to have $|\eta|<2.0$, whereas jets with a radius parameter $R = 0.4$  are referred to as {\textit{narrow}} jets, denoted as $j$ and are required to have $|\eta|<2.4$.
To mitigate the effects of pile-up and soft radiation, the large-$R$ jets are trimmed~\cite{jet_trimming}. Trimming takes the original constituents of the jet and reclusters them using the $k_t$ algorithm \cite{kt_algorithm} with a smaller radius parameter, $R_{\textrm{subjet}}$, to produce a collection of subjets. These subjets are discarded if they carry less than a specific fraction ($f_{\mathrm{cut}}$) of the original jet \pT. The trimming parameters optimised for this search are $R_{\textrm{subjet}} =0.2 $ and $f_{\mathrm{cut}}=5\%$~\cite{Aad:2015rpa}. Large-$R$ jets are calibrated following the procedure described in Ref.~\cite{ATLAS-CONF-2015-037}.

The energies of selected narrow jets are corrected for contributions from pile-up interactions~\cite{Aad:2015ina}. A correction used to calibrate jet energy measurements to the scale of the constituent particles of the jet~\cite{Aaboud:2017jcu} is then applied. Narrow jets with $25$~\GeV\xspace~$< p_{\textrm{T}} < 60$~\GeV\xspace are required to originate from the primary vertex as determined by a jet vertex tagger~\cite{Aad:2015ina} that relies on tracks associated with the jets.

Quality requirements are applied to photon candidates to identify those arising from instrumental problems or non-collision background~\cite{Aad:2014una}, and events containing such candidates are rejected. In addition, quality requirements are applied to remove events containing jets misreconstructed from detector noise or out-of-time energy deposits in the calorimeter from cosmic rays or other non-collision sources~\cite{jetsel}. 

\begin{figure}[t!]
    \centering
    \includegraphics[width=0.48\linewidth]{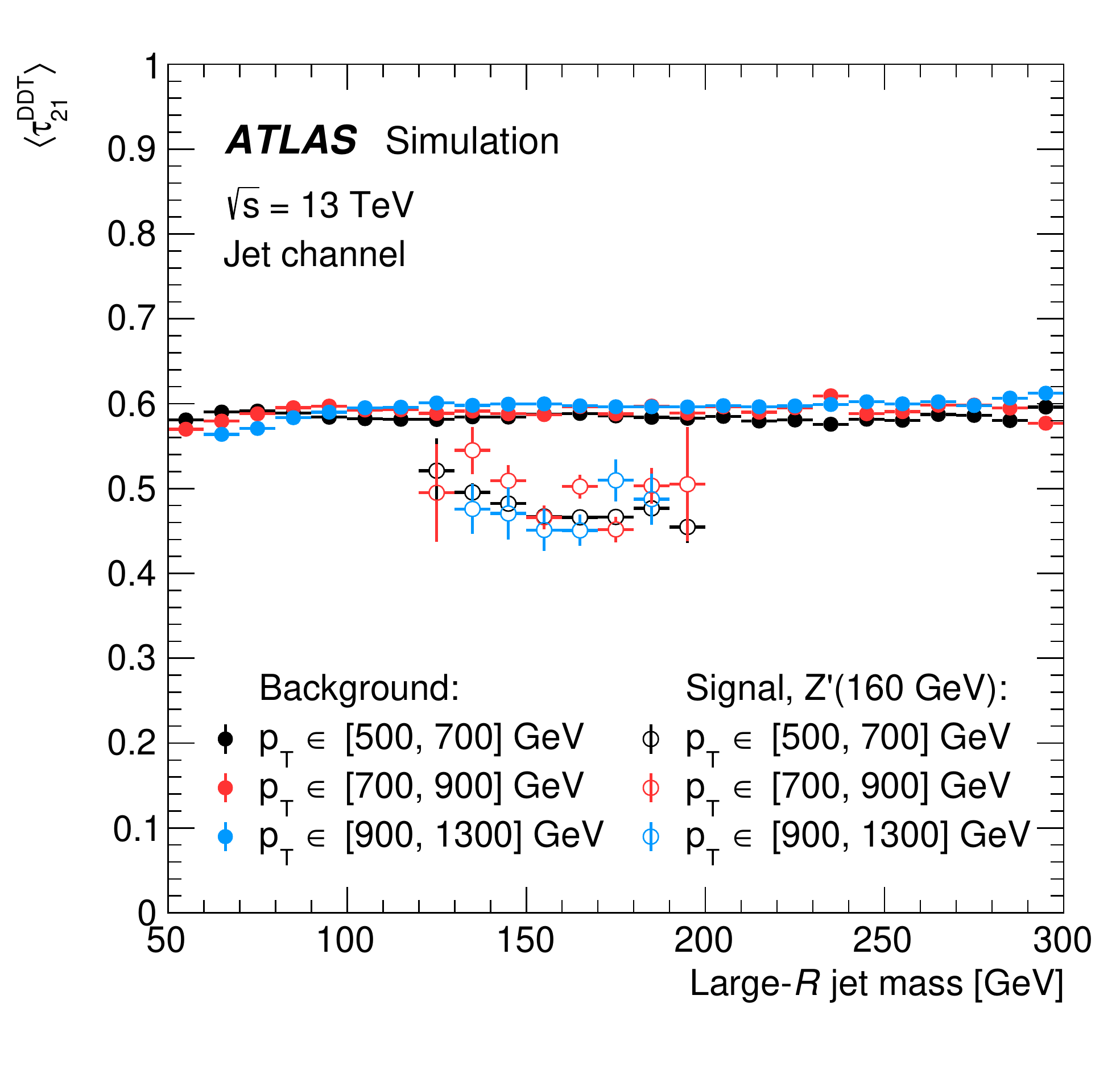}
    \includegraphics[width=0.48\linewidth]{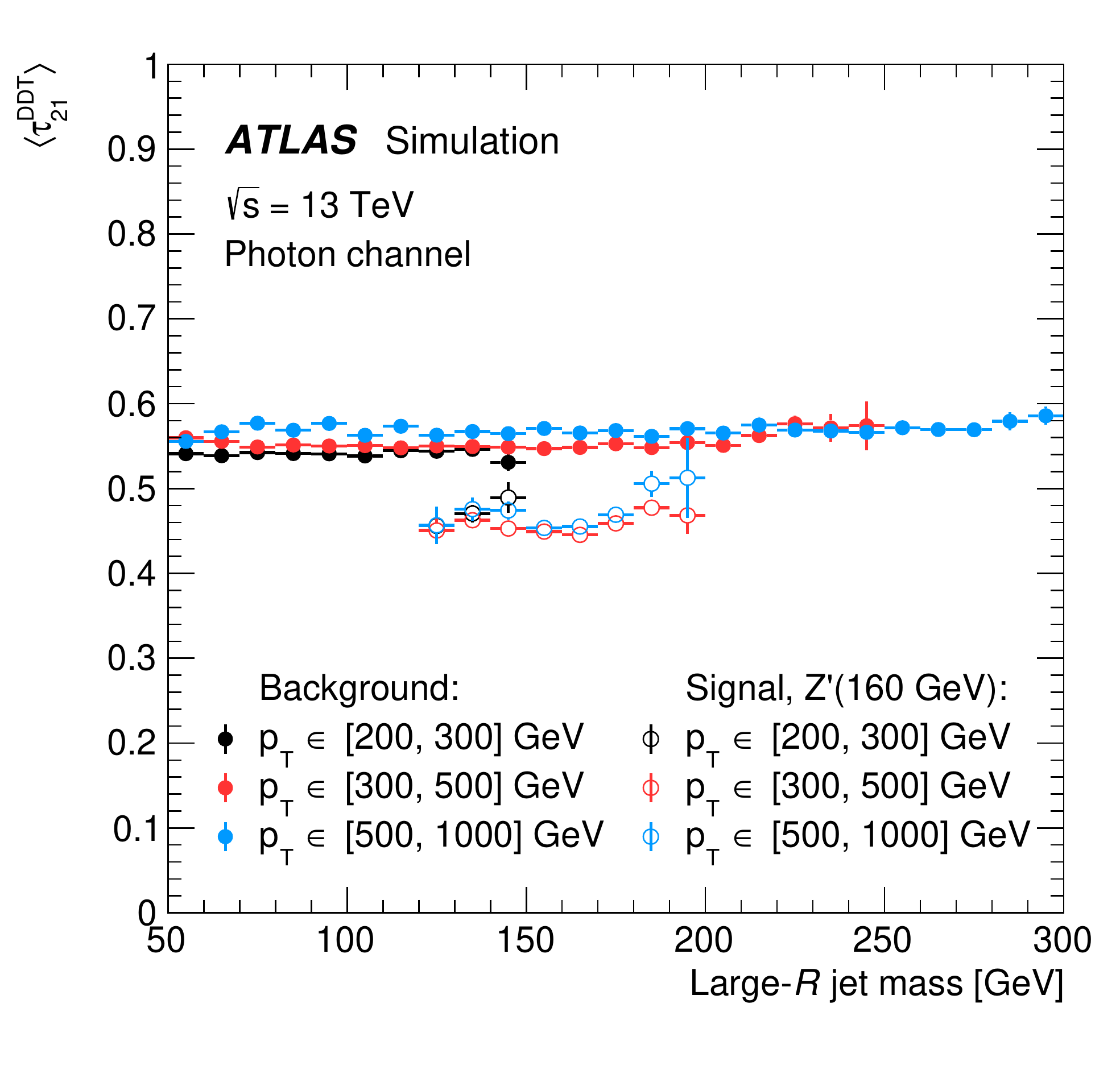}
    \caption{Mean value of $\tautwooneDDT$ as a function of the large-$R$ jet mass, for 
      various ranges of large-$R$ jet transverse momentum, for cases where the ISR object is a jet (left) and a photon (right). }
    \label{fig:ddt-tau21DDT}
\end{figure}

The production cross sections of the signal models considered in this search are many orders of magnitude lower than the background cross sections. 
In order to enhance the sensitivity to the signal, jet substructure techniques are used to identify the expected two-body quark-pair signal-like events within a single large-$R$ jet.
One of the commonly used jet substructure variables is $\tautwoone$~\cite{Thaler:2010tr}, 
defined as the ratio $\tau_2/\tau_1$. The variable $\tau_N$ is a measure of how consistent a given jet’s 
constituents are with being fully 
aligned along $N$ or more axes; thus $\tautwoone$ is a useful discriminant for 
differentiating between a two-particle jet from the decay of a boosted resonance and a single-particle jet.
However, $\tautwoone$ is correlated 
with the reconstructed large-$R$ jet mass $m_{J}$. 
Any selection requirement on $\tautwoone$ 
leads to a selection of jets from the leading background processes with efficiency strongly dependent on the jet mass,
and modifies the final jet mass distribution in a way that makes it difficult to model using a simple functional approach, effectively increasing the systematic uncertainties and weakening the overall sensitivity. To avoid this, the designed decorrelated tagger (DDT) method~\cite{Dasgupta:2013ihk,Dolen:2016kst,Sirunyan:2017dnz} is used to decorrelate $\tautwoone$ from the reconstructed jet mass. 
The variable \rhoDDT is defined as 
\[  \rhoDDT \equiv \log \left( \frac{m_{J}^{2}}{p^J_{\textrm{T}} \times \mu} \right),\]
where $\mu \equiv 1 ~\GeV$ is an arbitrary scale parameter.
For $\rhoDDT \gtrsim 1$, there is a linear relationship between $\rhoDDT$ and the mean value of $\tautwoone$.  
\rhoDDT is a purely kinematic jet variable, which allows 
the definition of $\tautwooneDDT$~\cite{Dasgupta:2013ihk,Dolen:2016kst}, a linearly corrected version of $\tautwoone$, which has mean values that are independent of the mass of the jet, as seen in Figure~\ref{fig:ddt-tau21DDT} for various ranges of large-$R$ $\ptJ$. 

Selected events are required to have at least one large-$R$ jet, the resonance candidate, and at least one narrow jet or photon with azimuthal angular separation of at least $\Delta \phi = \pi/2$ from the resonance candidate. 
The ISR jet is the leading narrow jet with $\ptj > 420 ~\gev$,  
while the ISR photon is the leading
photon with $\pty > 155 ~\gev$.  

In the signal region (SR), the large-$R$ jet must satisfy $\ptJ > 200 ~\gev$ in the photon channel and ${\ptJ > 450 ~\gev}$ in the jet channel.
Those thresholds are defined due to the minimum ${\ptJ > 200 ~\gev}$ for which large-R jets uncertainties have been derived (photon channel) and to select events with \ptJ close to the recoil jet \ptj, as expected for signal (jet channel). 
In addition, it is required that
$\ptJ > 2 \times m_{J}$ to ensure sufficient collimation of the quark pairs from signal resonances so as to avoid edge effects of using a fixed-cone jet algorithm, $\tautwooneDDT < 0.50$ to suppress backgrounds and $\rhoDDT > 1.5$. 
The $\tautwooneDDT$ requirement was chosen by maximising the expected signal significance.  
The $\rhoDDT$ constraint ensures that the $\tautwooneDDT$ variable is linear relative to $\rhoDDT$.
If multiple jets satisfy these requirements, the jet with the lower $\tautwooneDDT$ from the two leading  large-$R$ jets is selected. 

%-------------------------------------------------------------------------------
\section{Background estimation and systematic uncertainties}
\label{sec:background}
%-------------------------------------------------------------------------------

The dominant backgrounds in the jet and photon channels are due to multi-jet production and inclusive $\gamma$ production, respectively. 
The inclusive $\gamma$ background is dominated by $\gamma$+jets and also includes multi-jet processes being misidentified
with the same topology. 
In both channels, there is a sub-leading contribution from production of a jet or photon in association with a hadronically decaying electroweak gauge boson, $V$, where $V$ represents a $W$ or $Z$ boson.

In the dominant backgrounds, the boosted phase space relevant to this search is not well described by Monte Carlo programs.
Therefore, a data-driven technique is used to model the expected background in the signal region via a transfer-factor method which extrapolates from a control region (CR), defined by inverting the jet substructure requirement to \tautwooneDDT > 0.50.

The multi-jet and inclusive $\gamma$ background estimates are constructed in bins of candidate resonance mass. In each bin, the estimate is calculated as $(N_{\textrm{CR}} - N_{V})$ multiplied by the transfer factor, where $N_{\textrm{CR}}$ is the number of events in the CR and
$N_{V}$ is the expected contribution from production with an associated vector boson estimated from simulated samples, typically around 1\%. 
The transfer factor (TF) is the expected ratio of events which pass the \tautwooneDDT requirement to events which fail, measured using data with $m_{J} < (0.8 \times m_{Z'})$ or $m_{J} > (1.2 \times m_{Z'})$, to avoid potential contamination from a signal near $m_{Z'}$.  The TF is parameterised in terms of two kinematic quantities, $\log(\ptJ/\mu)$ and $\rhoDDT$; it is implemented as a two-dimensional histogram, smoothed and interpolated into the signal region using a Gaussian process (GP) regression~\cite{GaussianRegression} using a 
squared exponential or ``Gaussian kernel'' with a characteristic length scale $\ell \propto 1/\sigma$ for a Gaussian width $\sigma$.
The length scale $\ell_{d}$ along each dimension $d$ of the TF histogram in $(\rhoDDT, \log (\pT/\mu))$ is a free parameter, determined by maximising the marginal likelihood given by \cite{Lophaven:2002}:
\begin{equation*}
     \log L (\mathbf{y}\,|\,\mathbf{x},\{\ell_{d}\}) = - \frac{n}{2} \log\left[ \mathbf{y}^{\top} R_{\{\ell_{d}\}}(\mathbf{x}, \mathbf{x}) \, \mathbf{y} \right ] - \frac{1}{2}\log |R_{\{\ell_{d}\}}(\mathbf{x}, \mathbf{x})| 
\end{equation*}
where $\mathbf{x}$ and $\mathbf{y}$ are the measured TF histogram bins with values scaled to have zero mean and unit variance, $n$ is the number of data points, and $R_{\{\ell_{d}\}}(\mathbf{x}, \mathbf{x})$ is the correlation matrix of the TF measurements induced by the Gaussian kernel with length scales $\{\ell_{d}\}$. The TF values are regularised by the statistical uncertainties on the measurements according to Ref.~\cite{Lophaven:2002}. The first term quantifies the fit to the measurements, while the second term penalises model complexity (short length scales) \cite{GaussianRegression}.

The transfer factor is parametrised by ($\log(\ptJ/\mu)$, $\rhoDDT$) because \tautwooneDDT is decorrelated from $\rhoDDT$, making the transform factor maximally uniform along this variable. In addition, including $\log(\ptJ/\mu)$ in the parametrisation renders the dependence on the jet mass explicit, allowing for the construction of mass-dependent signal region windows.
The TF assume values between 0.6 and 1.3 across the ($\log(\ptJ/\mu)$, $\rhoDDT$) parameter space, in the jet channel, while the TF is between 0.5 and 0.9 in the photon channel. 
The difference in the TF distributions is due to the choice of the common \tautwooneDDT > 0.5 cut, which has comparable but not identical background acceptances in simulation for the two channels, while the spread in the range is due to discrepancies between data and simulation as well as the residual correlation between \tautwooneDDT and the jet kinematic parameters. 

Residual contamination from signal events which leak into the control region is accounted for in the statistical analysis   
as follows: the background estimate and its uncertainty are validated by constructing an 
interpolation  using data with $m_{J} < (0.7 \times m_{Z'})$ or $m_{J} > (1.3 \times m_{Z'})$, which is then compared to the data observed in a validation region (VR) in which $m_{J} \in [0.7,0.8] m_{Z'}$ or $m_{J} \in [1.2,1.3] m_{Z'}$. 
If the difference between the data and the background estimate in the VR is larger than the derived uncertainty, the uncertainty is inflated by a scale factor, without changing the nominal value of the background estimate. This can happen when the background estimate in the VR is derived from a control region with fewer events, and is therefore more sensitive to statistical fluctuations.
For the \ISRjet channel, the scale factor in the background uncertainty is found to be consistent with 1, while for the \ISRgamma channel the scale factor ranges from 1 to 2 across the values of $m_{Z'}$. This difference between channels comes from the number of events in data: the \ISRjet channel has 10 times more events than the \ISRgamma channel.

\begin{figure}[t!]
    \centering
    \includegraphics[width=0.48\linewidth]{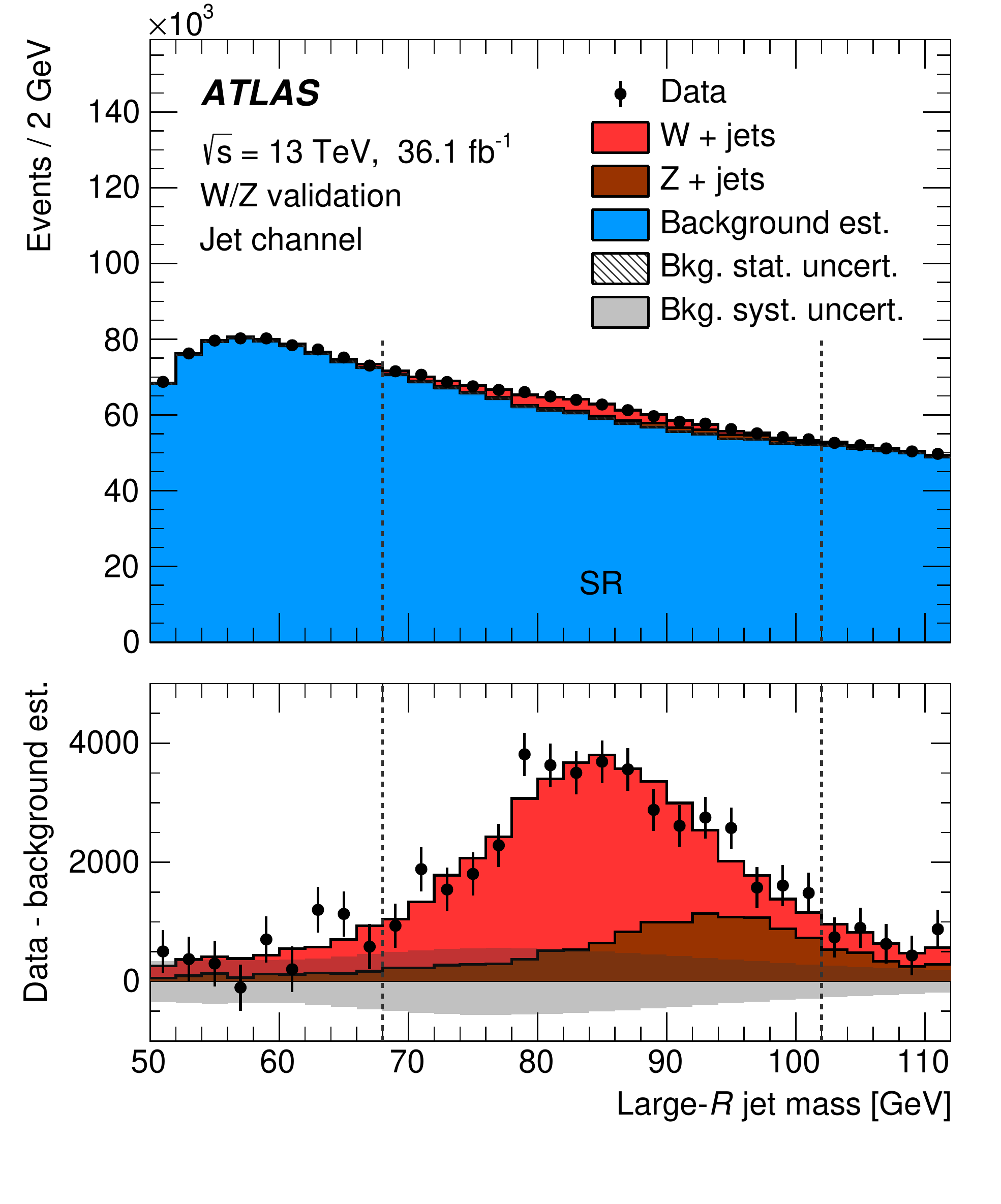}
    \includegraphics[width=0.48\linewidth]{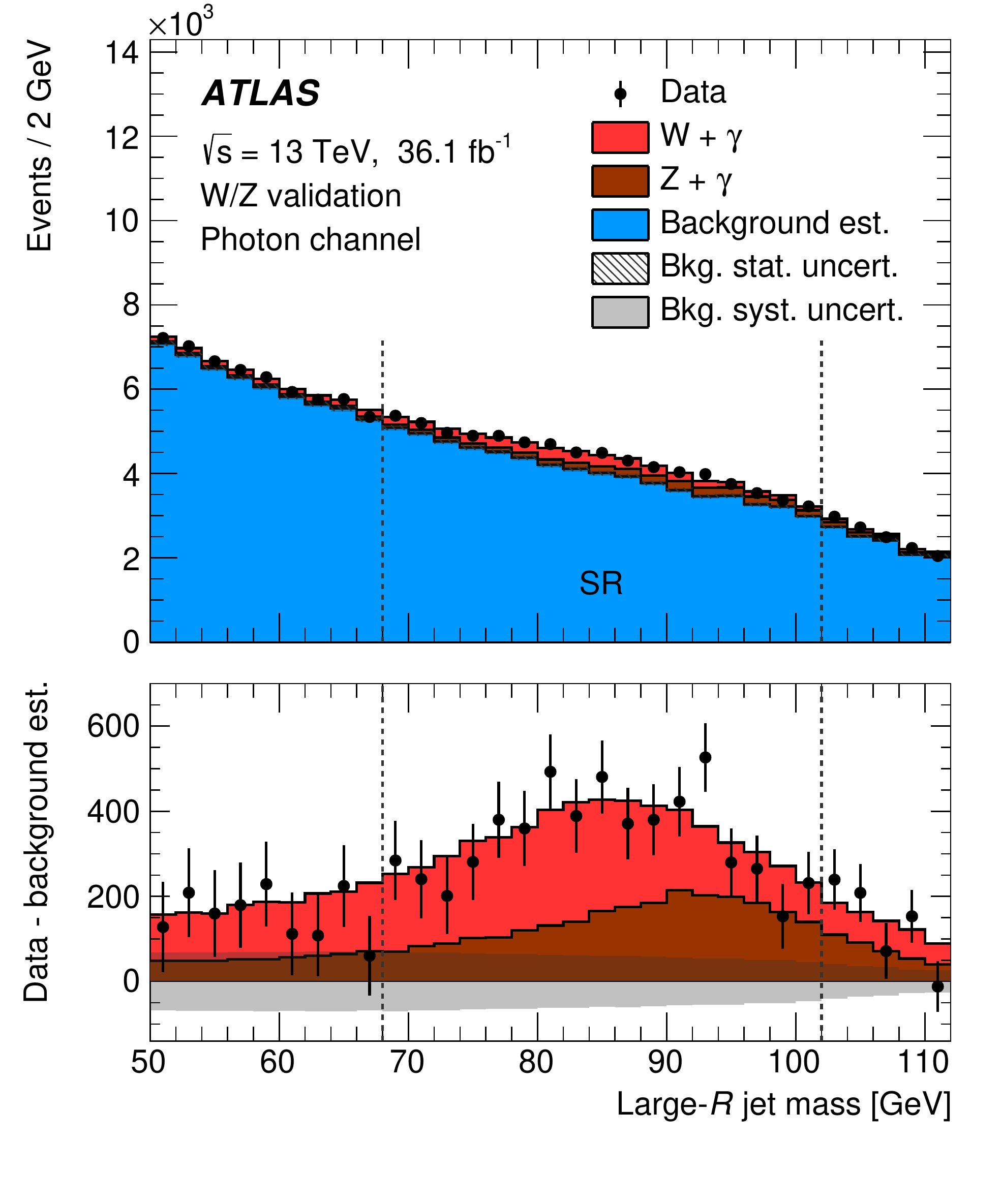}
    \caption{ Top: distribution of large-$R$ jet mass near the $W$ and $Z$ boson masses, as a validation of  background estimate using the transfer factor described in the text. The vertical dashed lines indicate the signal region (SR) surrounding the target $W$ and $Z$ boson masses.  
Bottom: residual between data and the estimated background.
The distributions are shown for both the (left) jet and (right) photon channels.  The contributions from the $W$ and $Z$ backgrounds have been scaled by their best-fit values, as described in the text.  In the top panel, the statistical uncertainty is too small to be visible; in the bottom panel it is incorporated into the error bars on the data. }
    \label{fig:WZ-20-mJ-postFit}
\end{figure}

As a cross-check, the TF method is applied to a candidate mass range near the $W$ and $Z$ boson masses:
the signal region's mass range is set as a $\pm 20$\% window around 85~\gev~([68,102]~\GeV), and the validation region as a $\pm 30$\% window around the same mass, but with the SR removed ([59.5,68]~\GeV~and [102,110.5]~\GeV).  Figure~\ref{fig:WZ-20-mJ-postFit} shows distributions of the large-$R$ jet mass for data 
and the resulting background estimate. The latter is found to agree with the data within uncertainties. The SM prediction for $W$ and $Z$ production is scaled with the NLO cross section using NLO $K$-factors, 
as described in Section~\ref{sec:samples}.
The cross sections used are 40.6 pb (18.6 pb) for the $W$($Z$)+jets processes in the \ISRjet channel, and 1.52 pb (0.983 pb) for $W$($Z$)+$\gamma$ processes in the \ISRgamma channel.
These cross sections are taken from the phase space of $p_T(W,Z)$ > 280 (140) GeV for the jet (photon) channels, as motivated by the analysis kinematic selections.
The best-fit signal strength relative to the SM prediction for $W$ and $Z$ production,
$\hat{\mu} = \sigma / \sigma_{W/Z}$, is \mbox{$\hat{\mu}$ = 0.93 $\pm$ 0.03 (stat) $\pm$ 0.24 (syst)} in the \ISRjet channel and 
\mbox{$\hat{\mu}$ = 1.07 $\pm$ 0.13 (stat) $\pm$ 0.35 (syst)} in the \ISRgamma channel, consistent with the SM predictions.
This result shows that the TF method works well.    

The largest systematic uncertainty is due to the estimate of the dominant background using the TF method. 
The Gaussian process regression provides a natural measure of the uncertainty in the interpolation, since it yields a mean function value across $(\log(\ptJ/\mu), \rhoDDT)$ and a covariance function $\mathrm{cov}(x,x’)$ relating the TF measurements at different $(\log(\pT/\mu), \rhoDDT)$. A $68\%$ confidence level uncertainty band, within which the true transfer factor is expected to lie~\cite{GaussianRegression}, can be obtained as $\sqrt{\mathrm{cov}(x,x)}$.
This uncertainty band, conditioned on the measurement of the ratio of numbers of events in the signal and control regions ($N_{\mathrm{SR}}/N_{\mathrm{CR}}$), is used as the systematic uncertainty on the transfer factor fit.
This uncertainty is tuned using the validation region defined above.
The final uncertainty is approximately 1\% of the total multi-jet or inclusive photon background estimate.

The uncertainty in the integrated luminosity is $2.1$\%; it is derived following a methodology similar to that detailed in Ref.~\cite{Aad:2013ucp}.
Additional systematic uncertainties stem from the use of simulated samples for the vector boson associated backgrounds as well as the hypothetical signals.
The largest sources of systematic uncertainty in each channel arise from uncertainties in the calibration and resolution of the large-$R$ jet energy and mass, as well as the modelling of $\tautwooneDDT$~\cite{ATLAS-CONF-2017-063}; individually these uncertainties range up to 10\% relative to the signal, but together these uncertainties are less than 1\% of the background estimate in the signal region.
Additional, smaller systematic uncertainties are due to the uncertainty in the parton distribution functions and integrated luminosity.

%-------------------------------------------------------------------------------
\section{Results}
\label{sec:results}
%-------------------------------------------------------------------------------

\begin{figure}[t!]
    \centering
    \includegraphics[width=0.48\linewidth]{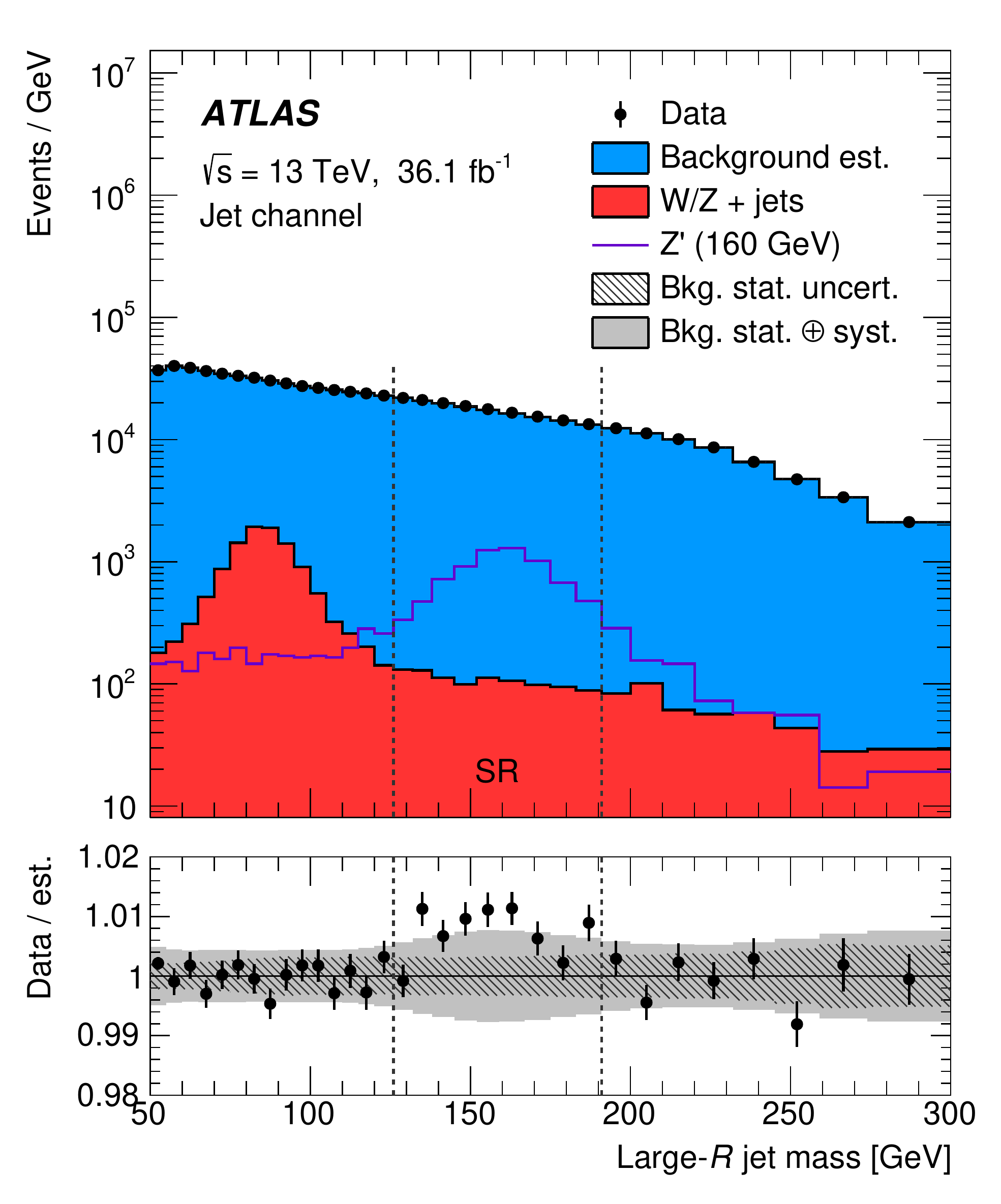}
    \includegraphics[width=0.48\linewidth]{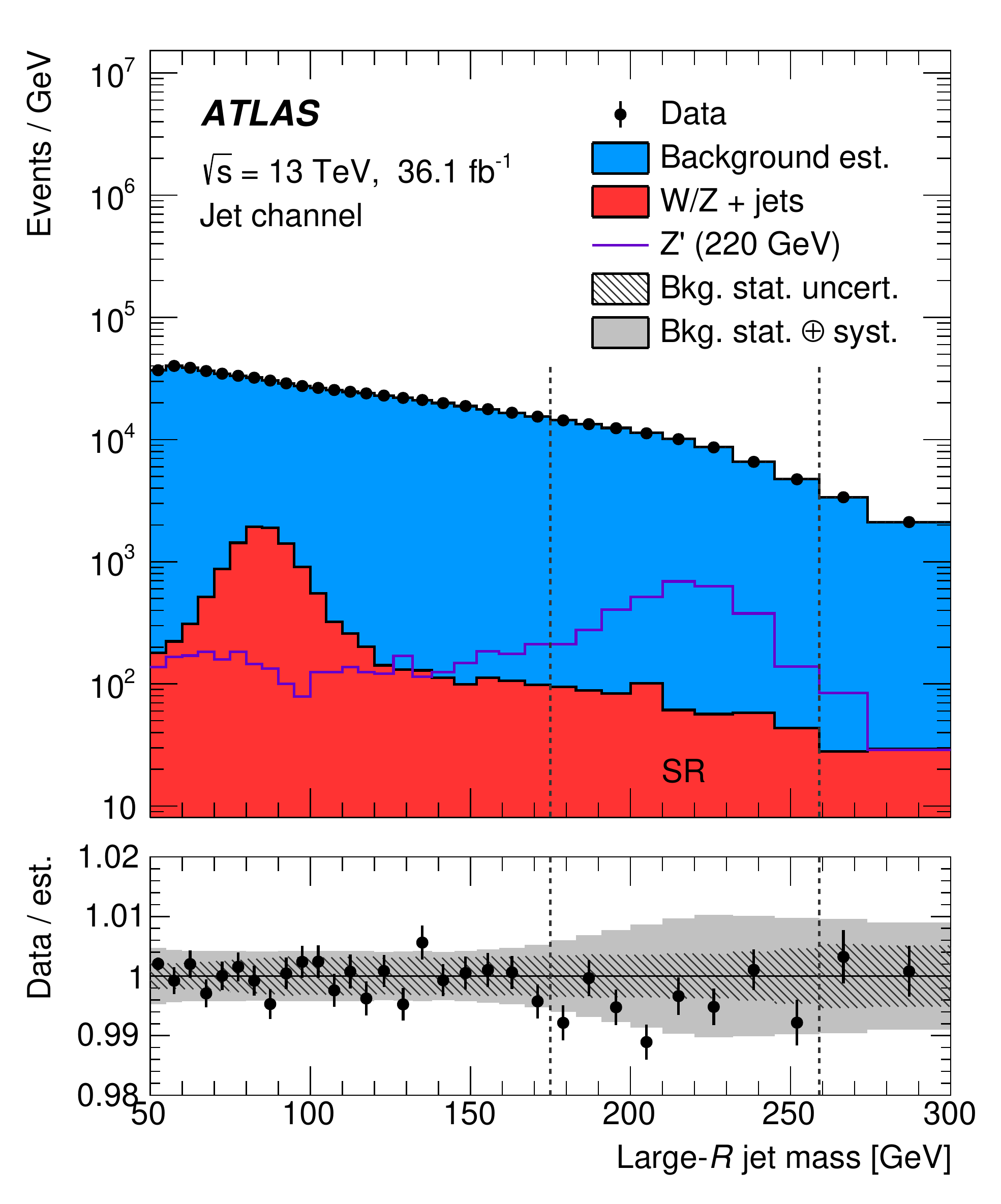}
    \caption{Top: distribution of large-$R$ jet mass in the jet channel for $m_{Z'}= 160 ~\gev$ (left) and $220 ~\gev$ (right).  The vertical dashed lines indicate the signal region (SR) surrounding the target $Z'$ mass.  
The signal is generated with $g_{q}$ = 0.5. 
Bottom: ratio of data to the estimated background.
The background estimate is different for each signal mass hypothesis; more details are given in the text.}
    \label{fig:unblinded-distributions-isrjet}
\end{figure}

\begin{figure}[t!]
    \centering
    \includegraphics[width=0.48\linewidth]{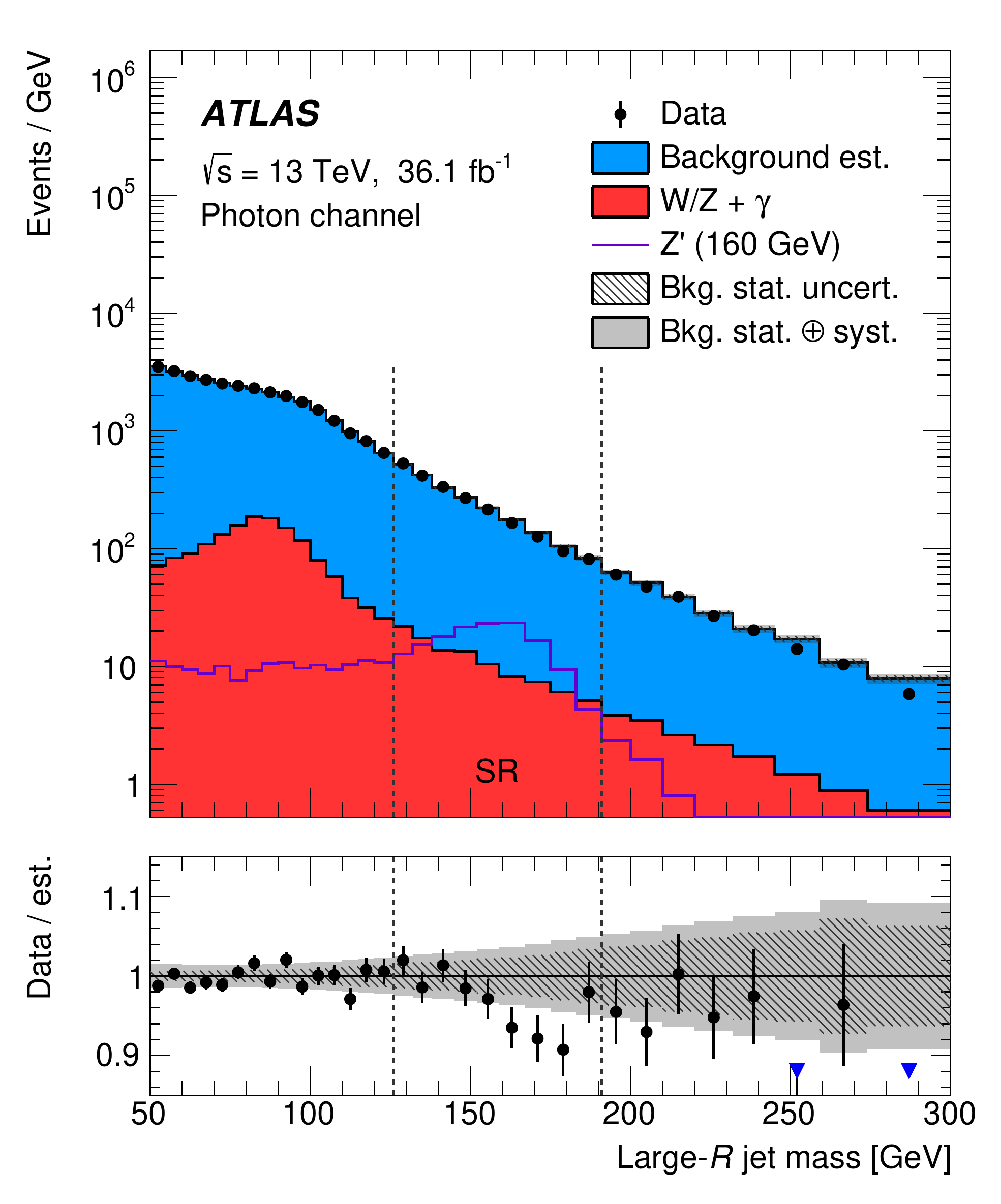}
    \includegraphics[width=0.48\linewidth]{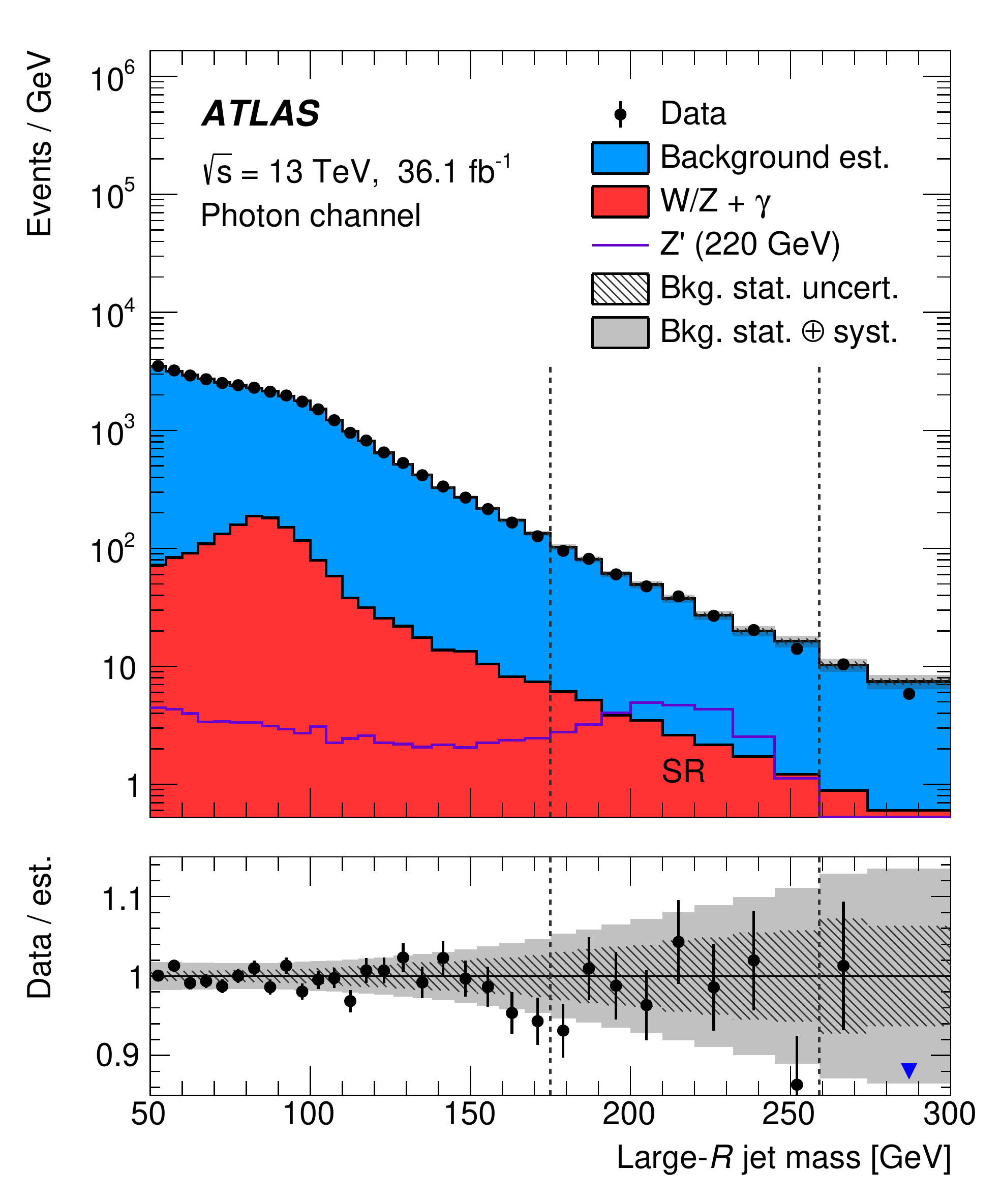}
    \caption{Top: distribution of large-$R$ jet mass in the photon channel for $m_{Z'}= 160 ~\gev$ (left) and $220 ~\gev$ (right). The vertical dashed lines indicate the signal region (SR) surrounding the target $Z'$ mass. The signal is generated with $g_{q}$ = 0.5. 
Bottom: ratio of data to the estimated background.
The background 
estimate is different for each signal mass hypothesis; more details are given in the text. The blue triangles indicate bins where the ratio is nonzero and outside the vertical range of the plot.}
    \label{fig:unblinded-distributions-isrgamma}
\end{figure}

The observed distributions of the large-$R$ jet mass are compared with the background estimates
in Figure~\ref{fig:unblinded-distributions-isrjet} and Figure~\ref{fig:unblinded-distributions-isrgamma}
for two representative $Z'$ mass values for the \ISRjet and \ISRgamma channels, respectively. 
The slope in the data and background distributions changes for a large-$R$ jet mass around $225 ~\GeV$ ($100 ~\GeV$) for Figure~\ref{fig:unblinded-distributions-isrjet} (Figure~\ref{fig:unblinded-distributions-isrgamma}), due to the boosted topology requirement,   
$\pT^{J} > 2 \times m_{J}$. The beginning of this effect is determined by the $\pT^{J}$ requirements of $450 ~\GeV$ and $200 ~\GeV$ for the \ISRjet and \ISRgamma channels, respectively.
The observed distributions of the large-$R$ jet mass are well reproduced by the estimated background contributions. 

A binned likelihood function $\mathcal{L}(\mu,\theta) $, constructed as a product of Poisson probability terms over all bins of the contributions of the background and of a hypothetical signal of strength $\mu$ relative to the benchmark model, is used to set limits. The likelihood function is also dependant on $\theta$, a set of nuisance parameters with Gaussian prior distributions encoding the effects of the systematic uncertainties in background and signal predictions. The fit to the large-R jet mass distribution is performed in each mass-dependent signal region in both the ISR jet and $\gamma$ channels. The potential signal contamination in the control region used to define the TF is accounted for by scaling the best-fit signal strength by the ratio of expected signal events passing the $\tautwooneDDT$ selection to the expected number of TF-weighted signal events included in the background estimation, as determined in simulation. Typical values for this scale factor are 0.7 for the \ISRjet channel and 0.6 for the \ISRgamma channel. 

The largest excess is observed in the \ISRjet signal region centred at 150~\GeV. Performing a signal-plus-background fit with a $Z'$ model assumption, 
the local significance in this region is found to be 2.5$\sigma$, corresponding to a global significance of 1.1$\sigma$, where the look-elsewhere effect~\cite{Gross:2010qma} is calculated with respect to the entire mass window examined.
The largest positive deviation from the expected background in the \ISRgamma channel is seen in the signal region centred at 
140~\GeV, with local (global) significance of 2.2$\sigma$ (0.8$\sigma$).

\begin{figure}[t!]
    \centering
    \includegraphics[width=0.48\textwidth]{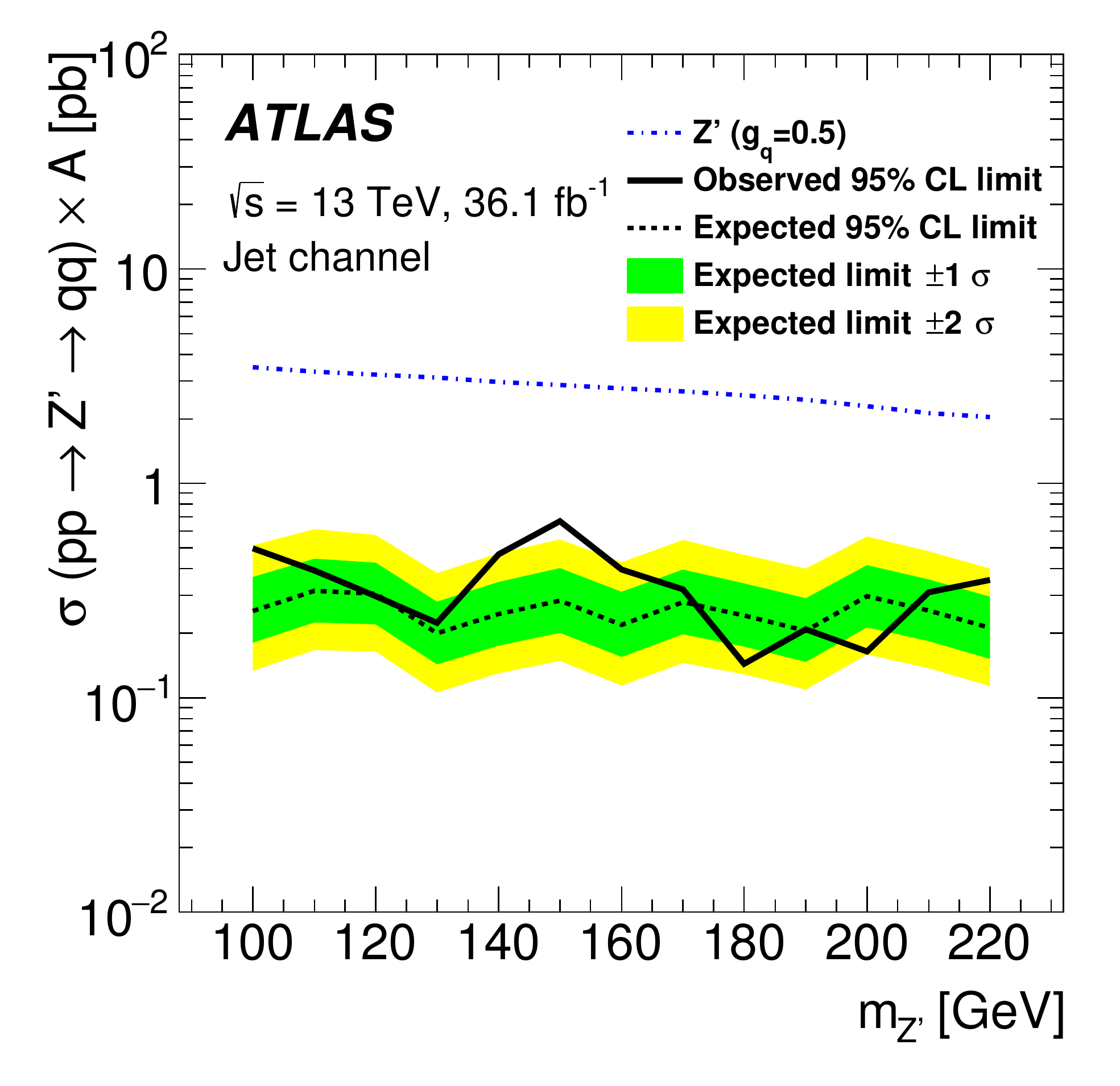}
    \includegraphics[width=0.48\textwidth]{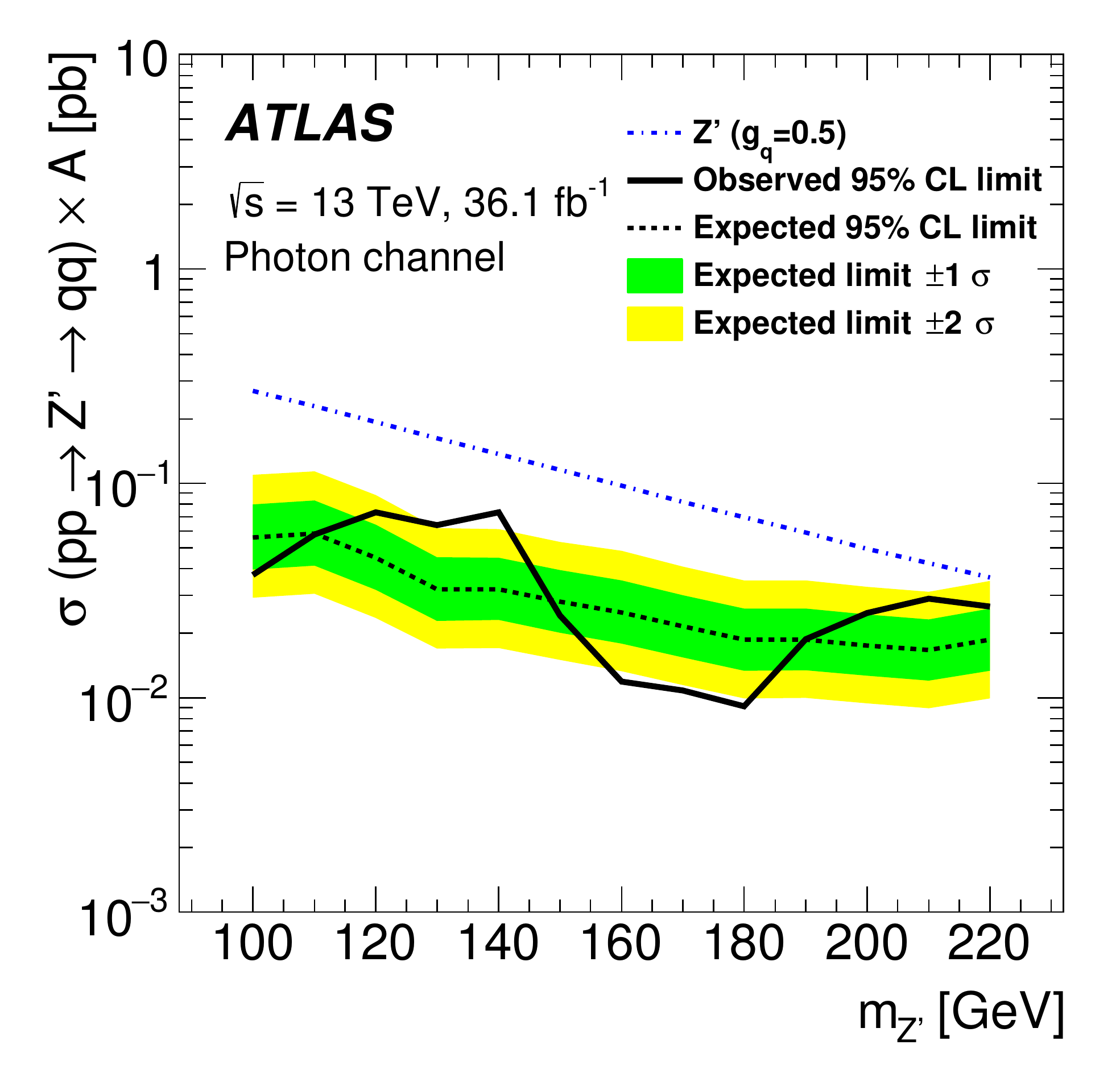}
    \caption{Observed and expected limits at 95\% confidence level on the lepto-phobic axial-vector $Z'$~\cite{Dobrescu:2013coa,LHCDMF:2015,Aad2016} production cross section ($\sigma$) times kinematic acceptance ($A$, see text for details) in the \ISRjet channel (left) and the \ISRgamma channel (right).}
    \label{fig:expectedLimitsDiJetISRjet}
\end{figure}

Upper limits are derived at 95\% confidence level on the $Z'$ production cross section  
times acceptance 
as a function of the $Z'$ mass between 
100 and 220~\GeV~using profile-likelihood-ratio tests~\cite{Cowan:2010js} with the CL$_{\mathrm{s}}$ method~\cite{cls}, shown in  
Figure~\ref{fig:expectedLimitsDiJetISRjet}. 

The acceptance accounts for all selection criteria except for the requirement on $\tautwooneDDT$; it can vary significantly for various theoretical models, yet can be well estimated without detailed detector simulation. 
For the $Z^{\prime}$ signal model considered in this paper, acceptance values vary from 0.10\% to 0.06\% in the \ISRjet channel and from 4.0\% to 1.0\% in the \ISRgamma channel, in the mass range between 100 and 220~\GeV.
The efficiency of the $\tautwooneDDT$ requirement is less model dependent but more dependent on accurate modelling of the $\tautwooneDDT$ 
variable in simulated samples. The acceptance times efficiency varies between 0.07\%--0.04\% (2.6\%--0.5\%) for the \ISRjet (\ISRgamma) channel over the 100--220~\gev~mass interval.

\begin{figure}[t!]
    \centering
    \includegraphics[width=0.48\textwidth]{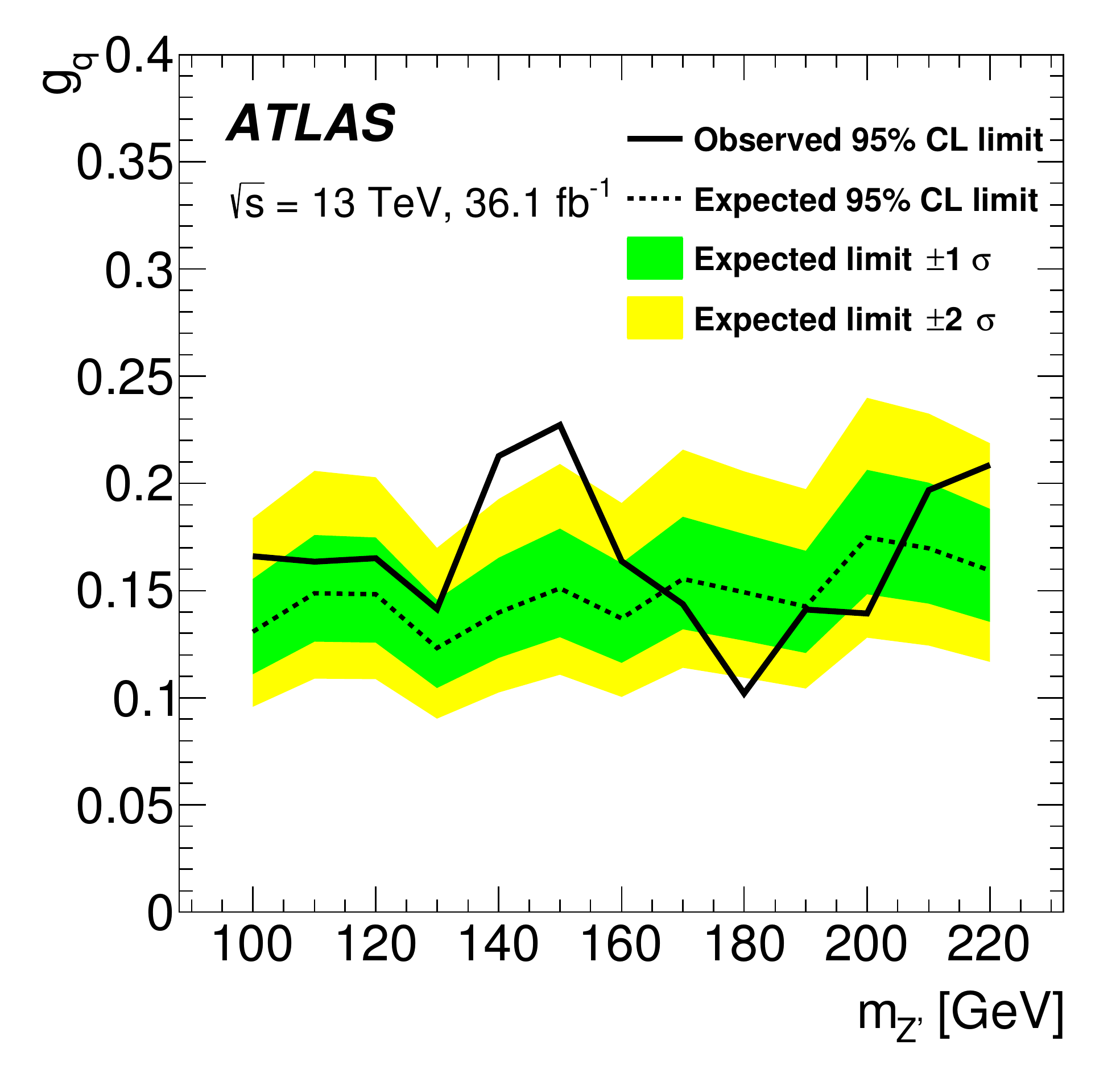}
    \caption{Observed and expected limits at 95\% confidence level on the coupling ($g_{q}$) from the lepto-phobic axial-vector $Z'$ model~\cite{Dobrescu:2013coa,LHCDMF:2015,Aad2016}, for the combination of the \ISRjet and \ISRgamma channels.}
        \label{fig:obsLimitsDiJetISRcoup}
\end{figure}

The observed and expected limits on the coupling $g_{q}$ are shown in Figure~\ref{fig:obsLimitsDiJetISRcoup}, 
for the combination of the \ISRjet and \ISRgamma channels. 
The narrow width approximation is valid for the $g_{q}$ range tested.
In the combination, the nuisance parameters corresponding
to luminosity and large-$R$ jet energy scale and resolution uncertainties are fully correlated between channels, while the background uncertainties are uncorrelated.
The largest deviation is for the 140~\GeV~signal hypothesis, corresponding to 2.4$\sigma$ local and 1.2$\sigma$ global 
significances. The observed upper limits on the coupling $g_{q}$ in the 100--220~\GeV~$Z^\prime$ mass range are competitive but slightly underperform the latest results reported by the CMS experiment~\cite{Sirunyan:2017nvi}, partially due to differences in the effect of jet trimming versus soft-drop grooming on relevant large-$R$ jet observables such as jet mass.

The effects of systematic uncertainties are studied for hypothesised signals using the signal-strength parameter $\mu$. The relative uncertainties in the best-fit $\mu$ value from the leading sources of systematic uncertainty are shown in Table~\ref{tab:impacts} for $m_{Z'}$ = 100, 160 and 220~\gev. The TF systematic uncertainty has the largest impact on the sensitivity, accounting for 86\%, 90\%  and 88\% of the total impact for the 100, 160 and 220~\gev~signal hypothesis, respectively.
The TF uncertainty is larger for the jet channel, due to its smaller length scale of the Gaussian process. For the $Z^\prime$ 160~\GeV~hypothesis, it accounts for 87\% of the impact in the signal strength in the \ISRjet channel and 61\% in the \ISRgamma channel.
The second biggest impact is due to uncertainties associated with large-$R$ jets. 
Ref.~\cite{ATLAS-CONF-2017-063} details the derivation procedure and the breakdown of those uncertainties.
The data's statistical uncertainty accounts for about 10\% of the total impact at all mass points considered. 
It is larger in the \ISRgamma channel than in the \ISRjet channel due to the order of magnitude difference in the number of events; this accounts for 21\% of the impact in the former and 9\% in the latter for $m_{Z^\prime}$ = 160~\GeV.

\begin{table}[t!]
\caption{
\label{tab:impacts}
The source of each of the largest uncertainties and their relative impact in the expected signal, quantified by the uncertainty in the best-fit signal strength ($\Delta \mu$) over the best-fit signal strength ($\mu$), for hypothesised signal production of $Z'$ with $m_{Z'}=100$~GeV, $m_{Z'}=160$~GeV and $m_{Z'}=220$~GeV. }
\centering
\begin{tabular}{lccc}
  \toprule
  Uncertainty source    & \multicolumn{3}{c}{$\Delta\mu/\mu$ [\%]}\\
   & $m_{Z'}=100$~GeV & $m_{Z'}=160$~GeV & $m_{Z'}=220$~GeV\\
  \midrule
  Transfer factor                     & 86  & 90                   & 88    \\
  Large-$R$ jet calib. and modelling  & 19  & 25                   & 17    \\
  W/Z normalisation                   & 43  &  $\ll$1                   &  $\ll$1    \\
  Signal PDF                          &  $\ll$1  &  $\ll$1                   &  1    \\
  Luminosity                          &  2  &  $\ll$1                   &  $\ll$1    \\
  Total systematic uncertainty        & 91  & 93                   & 91    \\
  \midrule
  Statistical uncertainty             &  9  & 10                   & 11    \\

  \bottomrule
\end{tabular}
\end{table}

\FloatBarrier

%-------------------------------------------------------------------------------
 \section{Conclusion}
 \label{sec:conclusion}
 %-------------------------------------------------------------------------------

In summary, a search for new light resonances decaying into pairs of quarks and produced in association with a high-$p_{\textrm{T}}$ photon or jet is presented. The search is based on 36.1 fb$^{-1}$ of 13 TeV $pp$ collisions recorded by the ATLAS detector at the LHC. Resonance candidates are identified as massive large-radius jets with substructure consistent with a quark pair. The   mass spectrum of the candidates is examined for local excesses above a data-derived estimate of a smoothly falling background. No evidence of anomalous phenomena is observed in the data, and limits are presented on the cross section and couplings of a leptophobic axial-vector $Z^\prime$ benchmark model. 
Upper limits at 95\% confidence level on production cross sections times acceptance are 0.50~pb (0.04~pb) for a 100~\GeV~signal hypothesis, 
and 0.35~pb (0.03~pb) for a 220~\GeV~signal hypothesis in the \ISRjet (\ISRgamma) channels. 
The observed upper limits on the coupling $g_{q}$ are 0.17 for $m_{Z^\prime}$ = 100~\gev~and 0.21 for $m_{Z^\prime}$ = 220~\gev, when combining \ISRjet and \ISRgamma channels.

%-------------------------------------------------------------------------------
\section*{Acknowledgements}
%-------------------------------------------------------------------------------

% Acknowledgements for papers with collision data
% Version 14-Feb-2018

% Standard acknowledgements start here
%----------------------------------------------
We thank CERN for the very successful operation of the LHC, as well as the
support staff from our institutions without whom ATLAS could not be
operated efficiently.

We acknowledge the support of ANPCyT, Argentina; YerPhI, Armenia; ARC, Australia; BMWFW and FWF, Austria; ANAS, Azerbaijan; SSTC, Belarus; CNPq and FAPESP, Brazil; NSERC, NRC and CFI, Canada; CERN; CONICYT, Chile; CAS, MOST and NSFC, China; COLCIENCIAS, Colombia; MSMT CR, MPO CR and VSC CR, Czech Republic; DNRF and DNSRC, Denmark; IN2P3-CNRS, CEA-DRF/IRFU, France; SRNSFG, Georgia; BMBF, HGF, and MPG, Germany; GSRT, Greece; RGC, Hong Kong SAR, China; ISF, I-CORE and Benoziyo Center, Israel; INFN, Italy; MEXT and JSPS, Japan; CNRST, Morocco; NWO, Netherlands; RCN, Norway; MNiSW and NCN, Poland; FCT, Portugal; MNE/IFA, Romania; MES of Russia and NRC KI, Russian Federation; JINR; MESTD, Serbia; MSSR, Slovakia; ARRS and MIZ\v{S}, Slovenia; DST/NRF, South Africa; MINECO, Spain; SRC and Wallenberg Foundation, Sweden; SERI, SNSF and Cantons of Bern and Geneva, Switzerland; MOST, Taiwan; TAEK, Turkey; STFC, United Kingdom; DOE and NSF, United States of America. In addition, individual groups and members have received support from BCKDF, the Canada Council, CANARIE, CRC, Compute Canada, FQRNT, and the Ontario Innovation Trust, Canada; EPLANET, ERC, ERDF, FP7, Horizon 2020 and Marie Sk{\l}odowska-Curie Actions, European Union; Investissements d'Avenir Labex and Idex, ANR, R{\'e}gion Auvergne and Fondation Partager le Savoir, France; DFG and AvH Foundation, Germany; Herakleitos, Thales and Aristeia programmes co-financed by EU-ESF and the Greek NSRF; BSF, GIF and Minerva, Israel; BRF, Norway; CERCA Programme Generalitat de Catalunya, Generalitat Valenciana, Spain; the Royal Society and Leverhulme Trust, United Kingdom.

The crucial computing support from all WLCG partners is acknowledged gratefully, in particular from CERN, the ATLAS Tier-1 facilities at TRIUMF (Canada), NDGF (Denmark, Norway, Sweden), CC-IN2P3 (France), KIT/GridKA (Germany), INFN-CNAF (Italy), NL-T1 (Netherlands), PIC (Spain), ASGC (Taiwan), RAL (UK) and BNL (USA), the Tier-2 facilities worldwide and large non-WLCG resource providers. Major contributors of computing resources are listed in Ref.~\cite{ATL-GEN-PUB-2016-002}.
%----------------------------------------------

%-------------------------------------------------------------------------------
%\clearpage
%\appendix
%\part*{Appendix}
%\addcontentsline{toc}{part}{Appendix}
%-------------------------------------------------------------------------------

%-------------------------------------------------------------------------------
% If you use biblatex and either biber or bibtex to process the bibliography
% just say \printbibliography here
\printbibliography
% If you want to use the traditional BibTeX you need to use the syntax below.
%\bibliographystyle{bib/bst/atlasBibStyleWoTitle}
%\bibliography{EXOT-2017-01-PAPER,bib/ATLAS,bib/CMS,bib/ConfNotes,bib/PubNotes}
%-------------------------------------------------------------------------------

\clearpage % ATLAS Collaboration author list
% Reference date of EXOT-2017-01 is 2017-07-31
% Author list last updated on date 06-AUG-18
% Data extracted on 06-Aug-2018 for paper reference EXOT-2017-01
% at 1:42pm
 
\begin{flushleft}
{\Large The ATLAS Collaboration}

\bigskip

M.~Aaboud$^\textrm{\scriptsize 34d}$,    
G.~Aad$^\textrm{\scriptsize 99}$,    
B.~Abbott$^\textrm{\scriptsize 124}$,    
O.~Abdinov$^\textrm{\scriptsize 13,*}$,    
B.~Abeloos$^\textrm{\scriptsize 128}$,    
S.H.~Abidi$^\textrm{\scriptsize 165}$,    
O.S.~AbouZeid$^\textrm{\scriptsize 143}$,    
N.L.~Abraham$^\textrm{\scriptsize 153}$,    
H.~Abramowicz$^\textrm{\scriptsize 159}$,    
H.~Abreu$^\textrm{\scriptsize 158}$
% ...
\bigskip
\\

% ...
$^{13}$Institute of Physics, Azerbaijan Academy of Sciences, Baku; Azerbaijan.\\
% ...
$^{34}$$^{(a)}$Facult\'e des Sciences Ain Chock, R\'eseau Universitaire de Physique des Hautes Energies - Universit\'e Hassan II, Casablanca;$^{(b)}$Centre National de l'Energie des Sciences Techniques Nucleaires (CNESTEN), Rabat;$^{(c)}$Facult\'e des Sciences Semlalia, Universit\'e Cadi Ayyad, LPHEA-Marrakech;$^{(d)}$Facult\'e des Sciences, Universit\'e Mohamed Premier and LPTPM, Oujda;$^{(e)}$Facult\'e des sciences, Universit\'e Mohammed V, Rabat; Morocco.\\
% ...
$^{99}$CPPM, Aix-Marseille Universit\'e, CNRS/IN2P3, Marseille; France.\\
% ...
$^{124}$Homer L. Dodge Department of Physics and Astronomy, University of Oklahoma, Norman OK; United States of America.\\
% ...
$^{128}$LAL, Universit\'e Paris-Sud, CNRS/IN2P3, Universit\'e Paris-Saclay, Orsay; France.\\
% ...
$^{143}$Santa Cruz Institute for Particle Physics, University of California Santa Cruz, Santa Cruz CA; United States of America.\\
% ...
$^{153}$Department of Physics and Astronomy, University of Sussex, Brighton; United Kingdom.\\
% ...
$^{158}$Department of Physics, Technion, Israel Institute of Technology, Haifa; Israel.\\
$^{159}$Raymond and Beverly Sackler School of Physics and Astronomy, Tel Aviv University, Tel Aviv; Israel.\\
% ...
$^{165}$Department of Physics, University of Toronto, Toronto ON; Canada.\\
% ...
$^{a}$ Also at  Department of Physics, University of Malaya, Kuala Lumpur; Malaysia.\\
$^{b}$ Also at Borough of Manhattan Community College, City University of New York, NY; United States of America.\\
$^{c}$ Also at Centre for High Performance Computing, CSIR Campus, Rosebank, Cape Town; South Africa.\\
$^{d}$ Also at CERN, Geneva; Switzerland.\\
$^{e}$ Also at CPPM, Aix-Marseille Universit\'e, CNRS/IN2P3, Marseille; France.\\
% ...
$^{*}$ Deceased

\end{flushleft}

% Created with Glance <Atlas.Glance@cern.ch>

%-------------------------------------------------------------------------------
% Auxiliary material - comment out the following line if you do not have any
%\include{EXOT-2017-01-PAPER-auxmat}
%-------------------------------------------------------------------------------

%-------------------------------------------------------------------------------
% Extra tables etc. for HepData - comment in the following line if you have any
% \include{EXOT-2017-01-PAPER-hepdata}
%-------------------------------------------------------------------------------

\end{document}